# Emergent normal fluid in the superconducting ground state of overdoped cuprates

Shusen Ye[1], Miao Xu[1], Hongtao Yan[2], Zi-Xiang Li[2,3,4], Changwei Zou[1], Xintong Li[1], Yiwen Chen[2], Xingjiang Zhou[2], Dung-Hai Lee[3,4] and Yayu Wang[1,5,6]*

[1]State Key Laboratory of Low Dimensional Quantum Physics, Department of Physics, Tsinghua University, Beijing 100084, P. R. China

[2]Beijing National Laboratory for Condensed Matter Physics, Institute of Physics, Chinese Academy of Sciences, Beijing 100190, P. R. China

[3]Department of Physics, University of California, Berkeley, CA, USA.

[4]Materials Sciences Division, Lawrence Berkeley National Laboratory, Berkeley, CA, USA.

[5]New Cornerstone Science Laboratory, Frontier Science Center for Quantum Information , Beijing 100084, P. R. China

[6]Hefei National Laboratory, Hefei, 230088, China

*Corresponding author. Email: yayuwang@tsinghua.edu.cn

**The microscopic mechanism for the disappearance of superconductivity in overdoped cuprates is still under heated debate. Here we use scanning tunneling spectroscopy to investigate the evolution of quasiparticle interference phenomenon in $Bi_2Sr_2CuO_{6+\delta}$ over a wide range of hole densities. We find that when the system enters the overdoped regime, a peculiar quasiparticle interference wavevector with quarter-circle pattern starts to emerge even at zero bias, and its intensity grows with increasing doping level. Its energy dispersion is incompatible with the octet model for *d*-wave superconductivity, but is highly consistent with the scattering interference of gapless normal carriers. The weight of the gapless quasiparticle interference is mainly located at the antinodes and is independent of temperature. We propose that the normal fluid emerges from the pair-breaking scattering between flat antinodal bands in the quantum ground state, which is the primary cause for the reduction of superfluid density and suppression of superconductivity in overdoped cuprates.**

# Main

One of the main puzzles concerning the cuprate high-temperature superconductors is the dome-shaped doping dependence of the critical temperature ($T_c$)[1]. The increase of $T_c$ in the underdoped regime is widely believed to be due to the increase of superfluid density hence phase stiffness by doping the parent Mott insulator, as illustrated by the Uemura law[2]. In the overdoped regime, the mechanism for the disappearance of superconductivity at the quantum superconductor to metal transition (QSMT) is still elusive[3]. It was usually considered to be due to the closing of superconducting gap like in conventional superconductors, but there are growing evidence that the gradual losing of phase coherence plays a crucial role in the QSMT. Especially, mutual inductance experiment reveals that overdoped cuprates exhibit an anomalous missing of superfluid density[4-6], which is intimately related to the decrease of $T_c$. The reduced phase stiffness results in strong phase fluctuation of order parameter and gradual gap filling across $T_c$, as demonstrated by angle-resolved photoemission (ARPES) experiments[7]. On the theoretical front, it has been proposed that the extensive pair-breaking scattering between the flat antinodal bands neighboring the van Hove singularity (vHS) is responsible for the suppression of phase stiffness[8-10]. When the disorder density is increased to $\frac{\hbar}{\tau} \geq \Delta_{SC}$, where $\Delta_{SC}$ is the superconducting gap and $\tau$ is the quasiparticle lifetime, pair-breaking scattering is expected to generate finite density of state (DOS) at the Fermi energy ($E_F$)[11-13]. Considering the conservation of spectral weight, the unpaired electrons would emerge from the extensive pair-breaking process and coexist with the superconducting condensate. The nature of such quantum normal fluid has yet to be demonstrated experimentally in detailed.

Scanning tunneling microscopy (STM) is a powerful technique to probe the normal fluid in the superconducting ground state owing to its ability to directly measure the single electron density of state (DOS) and image the characteristic scattering interference of normal carriers. Previous STM studies on overdoped cuprates have unveiled the restoration of a large hole-type Fermi surface[14-17], the closing of pseudogap[18], and the diminishing of checkerboard charge order[19,20]. Recently, the spatially periodic and particle-hole asymmetric modulation of the coherence peaks have been observed in overdoped $Bi_2Sr_2CaCu_2O_{8+\delta}$ (Bi-2212) and $Bi_2Sr_2Ca_2Cu_3O_{10+\delta}$ (Bi-2223), and is shown to be consistent with the strong pair-breaking antinode-antinode scattering induced by disorders[9]. However, due to the large antinodal $\Delta_{SC}$ in Bi-2212 and Bi-2223, the normal fluid generated by such pair-breaking has not been observed because the condition $\frac{\hbar}{\tau} \geq \Delta_{SC}$ is difficult to meet. In contrast, the single-layer compound $Bi_2Sr_2CuO_{6+\delta}$ (Bi-2201) has much smaller antinodal $\Delta_{SC}$ and widely tunable doping levels[21], providing an ideal system for investigating the emergent normal fluid and QSMT in the overdoped regime[3,8].

In this work, we use STM to investigate the evolution of quasiparticle interference (QPI) phenomenon in Bi-2201 over a wide range of hole densities. When the system enters the overdoped regime, a peculiar quarter-circle QPI wavevector starts to emerge even at zero bias. Its energy dispersion and temperature dependence are incompatible with the octet model for superconducting Bogoliubov quasiparticles with *d*-wave pairing symmetry, but are highly consistent with the scattering interference of gapless normal carriers. The emergent normal fluid in the superconducting ground state suggests that the sign-changing pair-breaking scattering between antinodal quasiparticles is the primary cause for the reduction of superfluid density and suppression of $T_c$ in overdoped cuprates[8].

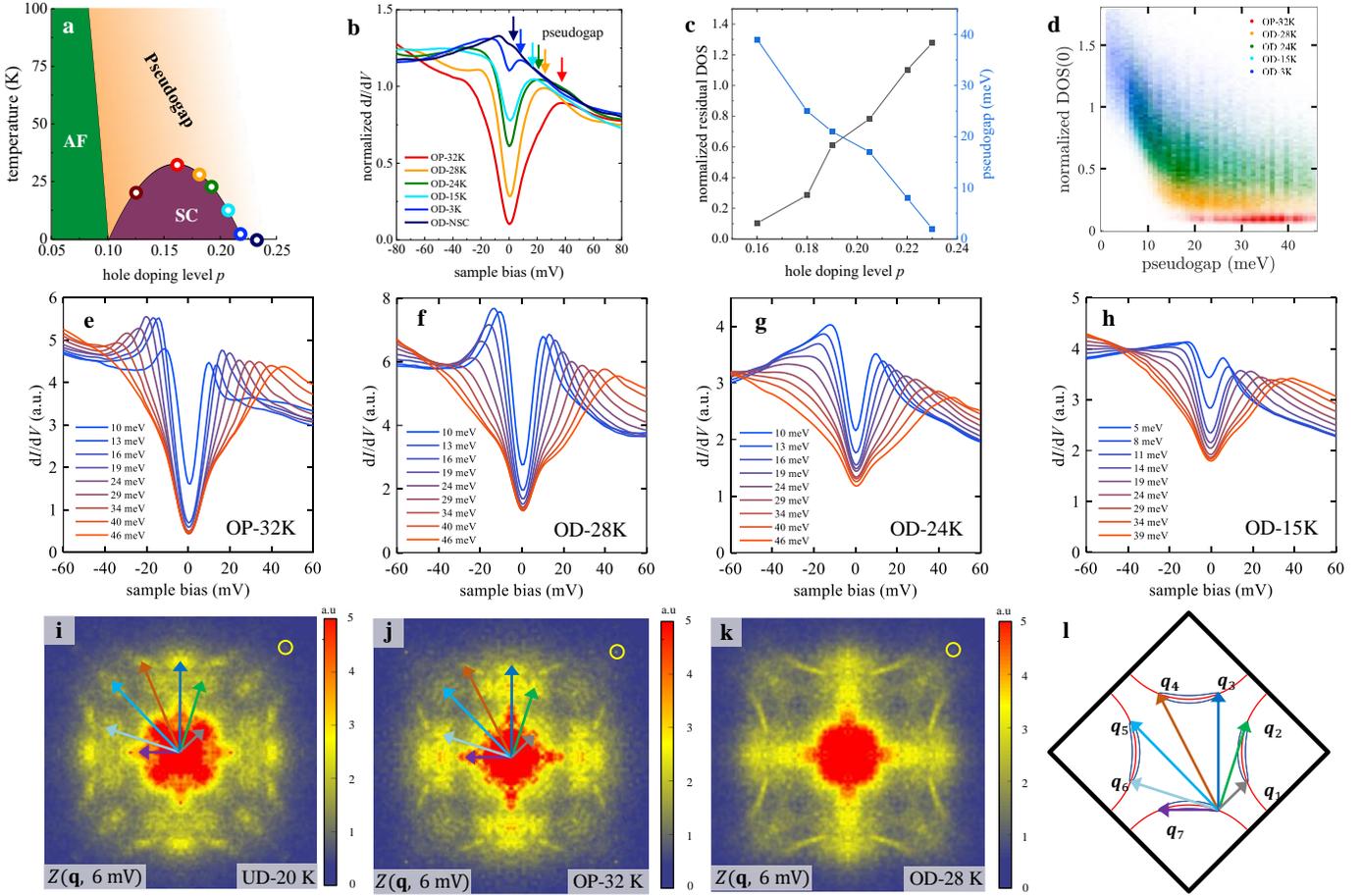

**Fig. 1| a**, The schematic phase diagram of Bi-2201. The open circles indicate the hole densities of the seven samples studied in this work. **b**, Spatially averaged d$I$/d$V$ spectra on the optimally doped and overdoped samples normalized by the mean DOS at ±70 mV. The pseudogap position of each curve is marked by an arrow with the corresponding color. The typical setup parameters are tunneling current $I$ = 150 pA and bias voltage $V$ = -100 mV. **c**, With increasing hole density in the overdoped regime, the pseudogap size decreases systematically while the normalized zero-bias DOS increases. **d**, Heatmap of the pseudogap and normalized zero-bias DOS extracted from the spatially resolved d$I$/d$V$ spectra over a large area in each overdoped superconducting sample. **e-h**, The averaged spectra sorted by the pseudogap size in the OP-32K, OD-28K, OD-24K and OD-15K samples. **i-k**, FT of the conductance ratio map $Z(\mathbf{q}, 6\text{ mV})$ on the UD-20K, OP-32K and OD-28K samples, respectively. The seven arrows indicate the dominant wavevectors, which are consistent with the octet model for Bogoliubov QPI with $d$-wave symmetry. **l**, Schematic diagram of the octet model with seven scattering wavevectors.

Figure 1a shows the schematic phase diagram of Bi-2201 with the circles indicating the seven samples studied in this work denoted as UD-20K, OP-32K, OD-28K, OD-24K, OD-15K, OD-3K and OD-NSC, respectively. Here UD/OP/OD represents underdoped/optimally doped/overdoped, the number represents $T_c$, and NSC represents non-superconducting. The hole densities are estimated to be $p$ = 0.13, 0.16, 0.18, 0.19, 0.21, 0.22 and 0.23, respectively, by the empirical $T_c$ versus $p$ relation[22] described in Supplemental Material Sec. I. Figure 1b displays the spatially averaged d$I$/d$V$ spectra normalized by the mean DOS values at ±70 meV, which demonstrates the evolution from the coexistence of superconducting gap (~10 meV) and pseudogap (~39 meV) in OP-32K[15,19] to a vHS peak in the OD-NSC sample[16]. In addition to the systematic decrease of pseudogap size $\Delta_{PG}$, another trend is the nearly linear increase of residual DOS at $E_F$ (or zero bias) with overdoping, as summarized in Fig. 1c. The increase of residual DOS at low temperature is consistent with the increasing residual electronic specific heat in overdoped cuprates[23-26]. To reveal the evolution of zero-bias DOS within each sample, we use clustering analysis to classify the large number of d$I$/d$V$ spectra obtained on a spatial grid by the pseudogap size, as shown in Figs. 1e-1h for the OP-32K, OD-28K, OD-24K and OD-15K samples. Within each sample, the zero-bias DOS always increases with reducing pseudogap size, hence the local hole density, and the trend becomes more pronounced in the more overdoped samples. The superconducting gap features around ±10 meV are evident in the large pseudogap clusters of each sample, and the gap size is nearly constant from the OP-32K to the OD-15K samples[27,28]. The heatmap in Fig. 1d directly visualizes the distribution of residual DOS versus $\Delta_{PG}$ in the five overdoped superconducting samples.

To elucidate the nature of low energy excitations, we obtain $g(\mathbf{r}, E)$ = d$I$/d$V(\mathbf{r}, E)$ over a large area in each sample, and use the Fourier transform (FT) map $g(\mathbf{q}, E)$ to identify the QPI wavevectors. To enhance the Bogoliubov QPI patterns due to superconductivity, we use the conductance ratio map $Z(\mathbf{r}, E) = g(\mathbf{r}, E)/g(\mathbf{r}, -E)$ to eliminate the influence of the setpoint effect[29]. The FT of Bogoliubov QPI patterns on the UD-20K and OP-32K samples are shown in Fig. 1e-f with seven arrows indicating the dominant wavevectors, which are consistent with the octet model for *d*-wave superconductivity as sketched in Fig. 1l[30]. Here the scattering mainly occurs between the hot spots of the banana-shaped equal-energy contours of the Bogoliubov quasiparticles[31-33]. With increasing doping level, the Bogoliubov QPI is gradually weakened, as reported by earlier studies on overdoped cuprates[34]. For the slightly overdoped OD-28K sample (Fig. 1g), the octet QPI wavevectors becomes nearly invisible.

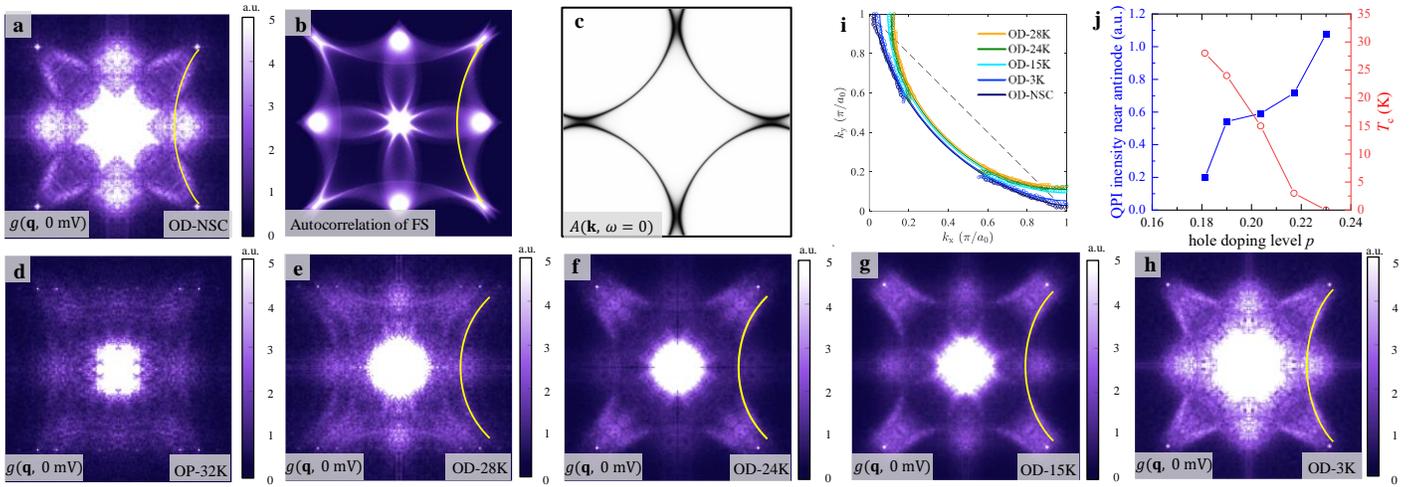

**Fig. 2 | a**, The zero-bias $g(\mathbf{q}, 0\text{ mV})$ map of the OD-NSC sample. The quarter-circle pattern reflects the QPI of hole-type Fermi surface. **b**, Autocorrelation of the Fermi surface calculated by the tight-binding model, which nicely reproduces the pattern in **a**. **c**, The hole-type spectral function at $E_F$ used in **b**. **d-h**, The zero-bias $g(\mathbf{q}, 0\text{ mV})$ maps on the OP-32K, OD-28K, OD-24K, OD-15K, and OD-3K samples, respectively. All the overdoped samples exhibit the quarter-circle feature highly resembling the normal state Fermi surface in the OD-NSC sample. **i**, The Fermi surface extracted from the $g(\mathbf{q}, 0\text{ mV})$ maps in the overdoped samples, revealing the increase of Luttinger volume with overdoping. **j**, The doping dependence of QPI intensity near antinodes in all overdoped samples with the respective $T_c$.

Next, we focus on the QPI phenomenon from the normal carriers by exploring the conductance map at zero bias. The Bogoliubov quasiparticles of a superconductor are gapped out at $E_F$, thus do not contribute to the zero-bias QPI. Therefore, the FT of zero-bias conductance map $g(\mathbf{q}, 0\ \text{mV})$ directly reveal the QPI originating from the normal carriers at $E_F$. The zero-bias QPI in the OD-NSC sample $g(\mathbf{q}, 0\ \text{mV})$ exhibits pronounced features consisting of quarter-circles highlighted by the yellow line (Fig. 2a). As described in a previous report[35], such patterns can be well reproduced by the autocorrelation[36] (Fig. 2b) of a single hole-like Fermi surface (Fig. 2c). As expected, in the OP-32K sample shown in Fig. 2d, the quarter-circle features are absent in $g(\mathbf{q}, 0\ \text{mV})$.

However, we find unexpectedly that even for the slightly overdoped OD-28K sample, the $g(\mathbf{q}, 0\ \text{mV})$ map shown in Fig. 2e already exhibits the quarter-circle QPI feature (the transversal replica is due to the structural supermodulation). With increasing doping (Figs. 2e-h), the zero-bias quarter-circle QPI features become more pronounced and the tip of the circle moves closer towards the vHS point. The striking resemblance to the OD-NSC sample and the smooth doping evolution strongly suggest that normal carriers start to emerge in the ground state when Bi-2201 enters the overdoped regime. In Fig. 2i we extract the Fermi surface of each sample following the same method for the OD-NSC sample[16], which displays a continuous expansion of the hole pocket with increasing $p$, providing further evidence for the normal carrier origin of the QPI. We then extract the QPI intensity near the antinode by integrating the $g(\mathbf{q}, 0\ \text{mV})$ along the Fermi wavevector at antinodal region (see Supplementary Material Sec. II for details), since the zero-bias QPI is mainly located at the antinodal region. As shown in Fig. 2j, the normal carrier QPI intensity grows with increasing doping and anti-correlates with $T_c$.

In fact, QPI with similar pattern has been observed in the conductance ratio map at finite energy[15], which is attributed to the antinodal Bogoliubov quasiparticle. However, the quarter-circle QPI of normal carrier is a different type of QPI, because it exists even at zero bias and does not exhibit the particle-hole symmetry about $E_F$. Because the Bogoliubov quasiparticle is generated by the opening of a superconducting gap at the normal carrier Fermi surface, they naturally share similar $q$-space pattern and may coexist at finite energy.

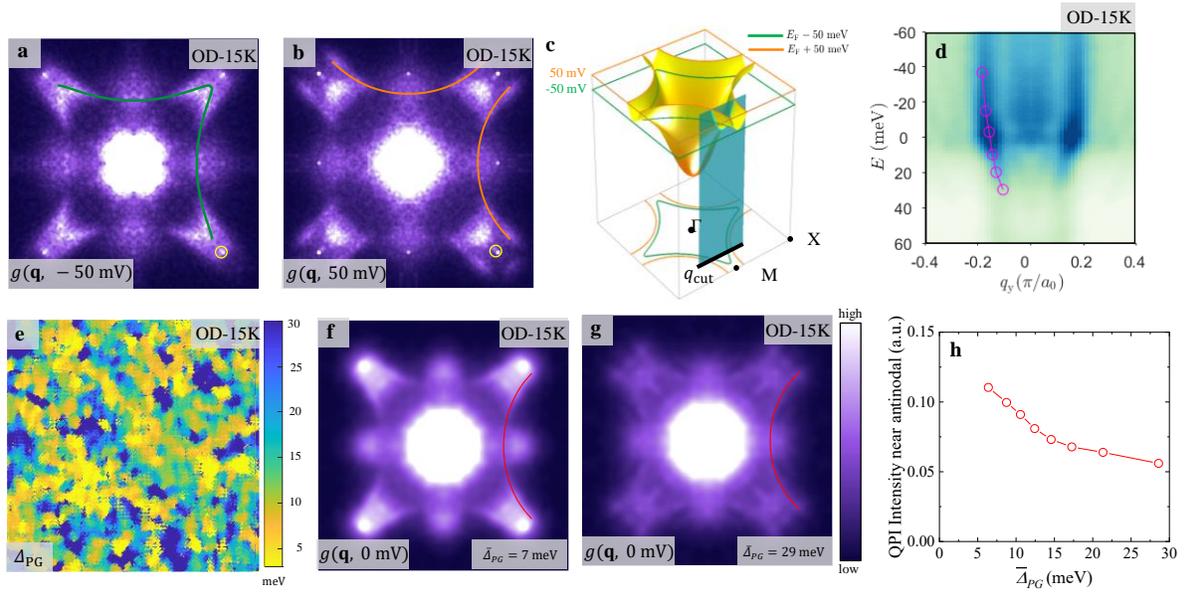

**Fig. 3 | a-b,** FT of the conductance maps $g(\mathbf{q}, -50\text{ mV})$ and $g(\mathbf{q}, 50\text{ mV})$ on the OD-15K sample. The green and orange lines are the equal-energy contours based on the tight-binding model. **c,** Schematic diagram of the band dispersion. The guidelines in **a-b** are sketched by the same colors and projected to the bottom plane. **d,** The intensity of QPI on the cross-section near the antinodal region indicated by the cyan plane in **c**, with open circles indicating the local maximum of QPI intensity. The dispersion demonstrates that the high-energy contours continuously extend to zero bias within the superconducting gap. **e,** The spatial distribution of pseudogap in the OD-15K sample shows typical nano-patch feature. **f-g,** The zero-bias QPI patterns in regions with averaged $\bar{\Delta}_{PG} = 7$ meV, and $\bar{\Delta}_{PG} = 29$ meV, respectively. **h,** The evolution of zero-bias QPI intensity near antinodes with the averaged local pseudogap size.

Figures 3a-b display the QPI patterns measured at $V = -50$ and $50$ mV on the OD-15K sample, corresponding to energies outside the pseudogap. The patterns match well with the equal-energy

contours of the normal carrier band structure obtained by the tight-binding model, as shown in Fig. 3c. For $V = -50$ mV, the band topology changes to that of a closed electron-type pocket near the antinodal region because now the energy lies below the vHS. To reveal the bias dependence of QPI, in Fig. 3d we plot the cross-sectional $g(q_y, E)$ with fixed $q_x = \frac{2}{3} Q_{Bragg}$, as illustrated by the cyan plane in Fig. 3c. It shows electron-type parabolic dispersion near the vHS at the M point, and the same plot for other overdoped samples exhibit systematic variations (see Supplementary Material Sec. III for details). More interestingly, it is apparent that the strongest QPI signals for all samples are located at $E_F$, which is strong evidence that they originate from normal carriers on the Fermi surface.

It may be argued that the normal carrier QPI in the superconducting ground state of overdoped cuprates comes from normal regions without superconductivity due to strong spatial inhomogeneity. To check if this is the case, we use the local pseudogap size as a proxy of local doping level, following a well-established practice[20]. The spatial distribution of $\Delta_{PG}$ in the OD-15K sample is displayed in Fig. 3e, which exhibits the usual nanoscale patches. We carry out clustering analysis of the whole field of view according the local pseudogap size, and obtain the FT of zero-bias QPI on each part (see Supplementary Material Sec. IV for details). The FT images within the smallest averaged pseudogap (7 meV) and the largest averaged pseudogap (29 meV) areas are displayed in Figs. 3f-g. The quarter-circle QPI exist in the $g(\mathbf{q}, 0 \text{ mV})$ of every regime with relatively weaker intensity in larger pseudogap area, as summarized in Fig. 3h, which is the same as the trend of different samples with varied doping levels (Fig. 2j). Interestingly, the larger $\Delta_{PG}$ regions show the quarter-circle QPI patterns with smaller Luttinger volume, confirming the validity of pseudogap size as local hole density marker. Therefore, the quarter-circle zero-bias QPI exists in all areas of an overdoped sample, and is not due to the phase-separated non-superconducting regions because of doping inhomogeneity. Besides, we have directly counted the percentage of normal regions in the OD-15K sample and find that it is merely ~10% (Supplementary Material Sec. V), thus is not enough to account for the strong quarter-circle zero-bias QPI. This is consistent with the finding of 12% gapless region in similar overdoped Bi-2201 with $T_c = 12$ K in another STM work using different data analysis method[37].

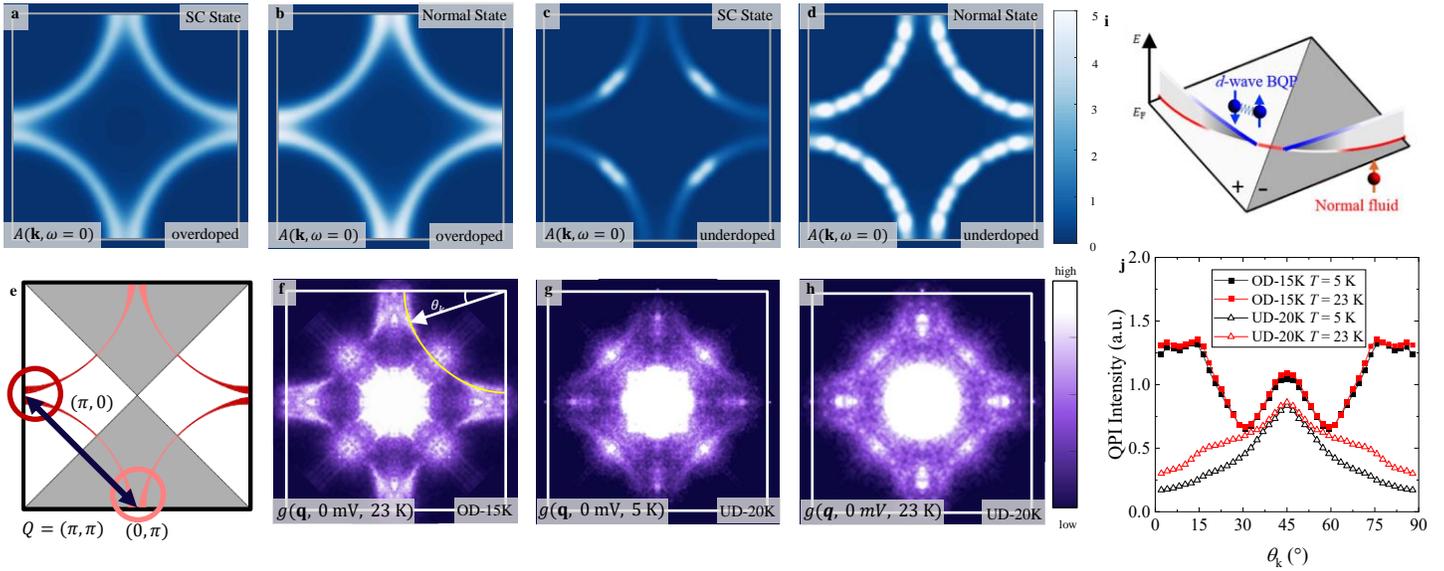

**Fig. 4** | **a-b**, The calculated spectral function $A(\mathbf{k}, \omega)$ at $E_F$ in the superconducting and normal states of overdoped cuprate. The $A(\mathbf{k}, \omega)$ at $E_F$ in the superconducting state is highly similar to the normal state Fermi surface. **c-d**, In the underdoped regime, the superconducting $A(\mathbf{k}, \omega)$ has residual excitation near the nodes with vanishing $\Delta_{SC}(\mathbf{k})$, while the normal $A(\mathbf{k}, \omega)$ displays a full Fermi surface. **e**, Schematic diagram of the pair-breaking scattering process with sign-changing superconducting order parameter between two antinodes. The vHS enhances such scattering due to the large joint DOS. **f**, FT of the conductance maps $g(\mathbf{q}, 0\,mV)$ in the OD-15K sample at 23 K, which is nearly identical to that at 5 K shown in Fig. 2g. **g-h**, FT of the conductance maps $g(\mathbf{q}, 0\,mV)$ in the UD-20K sample taken at 5 K and 23 K, respectively. The QPI pattern in the superconducting state is mainly located near the nodal region, while in the normal state higher intensity appears near the antinodal region due to thermally excited quasiparticles. **i**, Schematic diagram of the low-energy excitations in overdoped cuprates. In addition to that near the nodal region with

vanishing $\Delta_{SC}$, the pair-breaking scattering in the antinodal region creates a branch of normal fluid. **j**, Angle dependence of the zero-bias QPI intensity below and above $T_c$ for the UD-20K and OD-15K samples, revealing the strong contrast in the underdoped and overdoped regimes.

Although the existence of normal fluid in the superconducting ground state of overdoped cuprates has not been stressed in previous experimental reports, it is anticipated in a recent theoretical proposal[8]. The cross-antinode scattering is a strong pair-breaking process because of the sign-changing nature and the large DOS near the antinodal vHS, as shown in Fig. 4e. When the coherence length of disorder scattering $\xi_{scatter}$ is comparable to the superconducting coherence length $\xi_{SC}$, which is equivalent to the condition $\frac{\hbar}{\tau} \sim \Delta_{SC}$, gapless normal carriers appear and generate substantial unpaired electrons even in the superconducting ground state. To simulate the situations studied in this work, we use the self-consistent mean-field method to evaluate a $d$-wave superconductor with non-magnetic disorder scattering (see Supplementary Material Sec. VI for the details of simulation). As shown in Fig. 4a for the overdoped case, the approaching to the vHS generates large pair-breaking scattering rate, resulting in a nearly full Fermi surface even in the superconducting state, which is highly similar to that in the normal state (Fig. 4b). For the underdoped case, in contrast, the zero-energy electronic spectral function in the superconducting state mainly consists of nodal quasiparticles due to the vanishing of superconducting gap at the node (Fig. 4c). The antinodal spectral weight at Fermi energy only appears when the system enters the normal state (Fig. 4d).

To test the validity of this picture, we compare the zero-bias QPI on the same field of view of the OD-15K and UD-20K samples at $T$ = 5 K and 23 K, respectively, taken with the same experimental setup (see Supplementary Material Sec. VII for details of atomic scale tracking). For the OD-15K sample in Fig. 2g and Fig. 4f, the two sets of $g(\mathbf{q}, 0$ mV$)$ data are nearly identical, despite the fact that one is deep in the superconducting state while the other is above $T_c$. For the UD-20K sample, in contrast, there is an apparent enhancement of zero-bias QPI from below to above $T_c$ near the antinodal regime (Figs. 4g-h). The extracted angle dependence of QPI intensity in Fig. 4j reveals the thermally enhanced quasiparticles near the antinodal region in the UD-20K sample and the almost identical superconducting and normal state QPIs in the OD-15K sample. These results are highly consistent with the theoretical simulations (see Supplementary Material Sec. VIII for comparison with theoretical simulations), and demonstrate that sizable normal fluid density already exists in the superconducting ground state of overdoped cuprates (Fig. 4i).

The nearly temperature independent features also help distinguish the emergent normal fluid in overdoped cuprates from the classical two-fluid model[38], which originates from trivial thermal excitations and thus vanishes at low temperature. Moreover, the quantum normal fluid mainly emerges in the antinodal region with relatively large superconducting gap, which is also opposite to the expectation for thermally excited normal carriers or the finite instrument resolution because the core physics here is the enhanced pair-breaking for the flat antinodal band. We have performed similar experiment on overdoped Bi-2212 with $T_c = 66$ K, but find that the quarter-circle QPI pattern only appears at finite energy (see Supplementary Material Sec. IX). It is probably because the large antinodal $\Delta_{SC}$ makes it difficult to meet the condition $\frac{\hbar}{\tau} > \Delta_{sc}$ for generating sufficient normal carriers. This result is also consistent with the observation that low temperature residual electronic specific heat is absent in overdoped Bi-2212 with $p < 0.22$[39].

The zero-bias QPI patterns with quarter-circle wavevectors demonstrate the existence of normal fluid in the superconducting ground state of overdoped cuprates. The normal-carrier QPI features start to appear in the slightly overdoped OD-28K sample, where strongly overdoped normal regions are very scarce, indicating that the normal carrier QPI is not mainly due to real-space phase separation. It provides a natural explanation for the enhanced zero-bias DOS and diminishing Bogoliubov QPI with overdoping as shown in Fig. 1, as well as the low temperature residual electronic specific heat[23-26]. It can also explain the anomalous missing of superconducting electrons observed by mutual inductance[4], the two-component fluid transport behavior[40], and the uncondensed Drude-like peak at low temperature in overdoped cuprates[5]. Moreover, it is consistent with the postulation that the Planckian quadrature magneto-resistance in overdoped cuprates is due to the discontinuity of Fermi surface sections induced by the $k$-space distribution of normal carriers[40-43]. The recent finding of the extension of superconducting dome to more overdoped regime in a low-disorder cuprate system is also consistent with this picture[44]. Most importantly, our results provide a microscopic mechanism for the suppression of superconductivity in the overdoped regime. The cross antinodal pair-breaking scatterings are responsible for the reduction of superfluid density and suppression of $T_c$, which eventually drive the QSMT[8].

## Methods

**Sample growth.** High-quality Bi-2201 single crystals with La or Pb substitutions are grown by the traveling solvent floating zone method and annealed in $O_2$ under different conditions to tune the hole density over a wide range, as described in a previous report[21]. The chemical formulas of the seven samples can be expressed as $(Bi_{2-x},Pb_x)(Sr_{2-y},La_y)CuO_{6+\delta}$. The detailed ingredients are described in Tab. S1.

**STM measurements.** The single crystals are cleaved *in situ* at 77 K in the ultrahigh vacuum preparation chamber and immediately transferred into the STM stage cooled to $T = 5$ K. The STM topography is collected in the constant current mode with an etched tungsten tip, which is calibrated on a clean Au(111) surface. The differential conductance d$I$/d$V$ is acquired by a standard lock-in technique with frequency $f$ = 723.137 Hz. The data supporting the analysis in the main text contains a series of spectral grids on samples with different doping levels, and the detailed parameters are listed in Tab. S2. Before the FT analysis of QPI data, the Lawler-Fujita algorithm[45] is applied on the d$I$/d$V$ map to guarantee the accuracy of momentum coordinate. The FT of QPI data are symmetrized along the high-symmetry axis of lattice to enhance the signal-to-noise ratio.

## Data availability

All data are available in the main text or the supplementary materials.

## Acknowledgments

We thank Z.Y. Weng and G.M. Zhang for helpful discussions. This work was supported by the Basic Science Center Project of NSFC under grant No. 52388201, NSFC Grant No. 11888101. Y.W. is supported by the Innovation Program for Quantum Science and Technology (grant No. 2021ZD0302502), and the New Cornerstone Science Foundation through the New Cornerstone Investigator Program and the XPLORER PRIZE. X.J.Z. is supported by the Strategic Priority Research Program (B) of the Chinese Academy of Sciences (XDB25000000).

## Author contributions

Y. W. and X. Z. supervised this project. H. Y. and Y. C. prepared the single crystal. S. Y., C. Z. and M. X. carried out the STM experiments under the supervision of Y.W. Z. H., Y. J., and S. Y. performed the transport experiments. S.Y. and Y.W. prepared the manuscript with comments from all authors.

## Competing interests

The authors declare no competing interests.